\documentclass[graybox]{svmult}

\usepackage{mathptmx}       
\usepackage{helvet}         
\usepackage{courier}        
\usepackage{type1cm}        
\usepackage{makeidx}         
\usepackage{graphicx}        
\usepackage{multicol}        
\usepackage{url}        
\usepackage[bottom]{footmisc}

\makeindex             

\begin{document}

\title*{Bayesian mixture models for Poisson astronomical images}
\author{F. Guglielmetti$^{1}$ \and R. Fischer$^{2}$ \and V. Dose$^{2}$}
\institute{Fabrizia Guglielmetti 
\at Max-Planck-Institut f\"ur extraterrestrische Physik, Giessenbachstrasse, D-85748 Garching, Germany, \email{fabrizia@mpe.mpg.de}
\and Rainer Fischer \and Volker Dose 
\at Max-Planck-Institut f\"ur Plasmaphysik, Boltzmannstrasse 2, D-85748 Garching, Germany, \email{Rainer.Fischer@ipp.mpg.de, vod@rzg.mpg.de}}
\maketitle
\abstract{Astronomical images in the Poisson regime are typically characterized by a spatially 
varying cosmic background, large variety of source morphologies and 
intensities, data incompleteness, steep gradients in the data, and few photon counts per 
pixel.
The Background-Source separation technique is developed with the aim 
to detect faint and extended sources in astronomical images characterized by Poisson statistics. 
The technique employes Bayesian mixture models to  
reliably detect the background as well as the sources with their respective uncertainties. 
Background estimation and source detection is achieved in a single algorithm.  
A large variety of source morphologies is revealed. 
The technique is applied in the \emph{X}-ray part of the electromagnetic spectrum 
on \emph{ROSAT} and \emph{Chandra} data sets and it 
is under a feasibility study for the forthcoming \emph{eROSITA} mission.
}

\section{Introduction}\label{sec:1}
One of the hot topics in \emph{X}-ray (quantum energies $>0.1$ keV) image analysis is the 
detection of faint sources. Both point-like and extended faint sources may provide important 
information about the Cosmos. 
For instance, a quantitative analysis of the abundance of galaxy clusters and groups as a function 
of redshift allows one to constrain cosmological parameters, to test the models for structure 
formation and to provide the basis for follow-up studies of physical properties of these systems 
\cite{rosati-araa}, \cite{boehringer:2010}.
The detection and characterization of faint sources require advanced statistical methods. 

\subsection{The data}
\emph{X}-ray images are characterized by few or no photon counts per pixel also for long 
exposures. 
The data consists of a diffuse background with superposed celestial objects, 
corrupted by Poisson noise and affected by instrumental complexities. 
Poisson noise dominates the signal especially at high 
frequencies of the electromagnetic spectrum. 
The instrumental complexities are, e.g., exposure variations, 
instrumental structures as detector ribs 
and charge-coupled device (CCD) gaps, smearing and vignetting 
effects, CCD failures and instrumental calibrations. 
An astronomical image is often a combination of several individual 
pointings, as for deep observations and mosaics of images, and the effects due to 
steep gradients in the data are cumbersome. Furthermore, the \emph{X}-ray background is a 
composition of instrumental, particle and cosmic emissions. 
The cosmic background is not necessarily spatially constant. 
Celestial objects are characterized by a large variety of 
morphologies and apparent brightnesses.
Sources, especially extended ones, can be superposed to both, smooth and highly, varying 
background.

\subsection{Challenges in image analysis}\label{sec:DandC}
The interpretation of observational data is a difficult task, especially when detecting faint sources and 
their (complex) morphologies. Several approaches have been developed so far. However, previous techniques do not 
jointly detect a large variety of source morphologies and describe large variations in the 
background.  

An ideal source detection method should be capable to, preserve the statistics through the whole 
algorithm, detect faint sources, detect both point-like and extended sources, including complex 
morphologies, provide an accurate background estimation, include the exposure map in the background 
model, and provide uncertainties of estimates. 
Each of these desiderata entail a challenge in source detection and background estimation. 
In fact, the nature of the data of \emph{X}-ray images is described by Poisson statistics and 
Poisson noise affects the data.
Furthermore, joint background estimation and source detection is essential for a reliable 
detection of celestial objects and for a proper propagation of errors in background and source 
estimates. 
Conventional methods employ a threshold level for separating the sources from the 
background. 
Often, the threshold level is 
described in terms of the noise standard deviation, then translated into a probability 
(\emph{p}-values). An ideal source detection method has to replace the threshold level by a measure 
of probability. 
In the same line of arguments, parameters entering the models need to be estimated from the data. 
In addition, the detection of extended sources is commonly achieved in several steps, e.g., reanalyzing 
the image after removing point sources from the image. Consequently, uncertainties in the data are not 
properly accounted for.
In order to detect faint and extended sources, source features extending to the edge of the field of view 
and for providing good estimates in object photometry, a stable background model is essential.  
The estimation of a reliable background model and its uncertainties is a demanding task. 
Many techniques subtract an estimated background from the data, 
leading even to negative count rate values of the signal of interest: See, e.g., \cite{sliwa:2001}. 
Moreover, the background model has to incorporate the exposure map. 
Exposure maps include also factors such as vignetting, defective pixels and instrumental structures, 
resulting in lack of data. The missing data must be handled 
consistently for the background estimation to prevent undesired artificial effects. 
Hence, the challenge is to preserve the statistics 
while taking into account the exposure map in the background model. 
The last demanding aspect for an ideal source detection method is the proper quantification of uncertainties 
of estimates. 

Note that the knowledge of the instrumental point-spread-function (PSF) is not considered 
essential for source detection. A source detection algorithm designed for the detection of 
a large variety of source morphologies should be able to operate effectively without the PSF information.
Source detection methods employing a PSF or its functional form are designed for the
detection of point-like objects regardless of extended ones \cite{starck:1998}. 

\section{The Background-Source separation algorithm} \label{sec:BSS}
The Background-Source separation (BSS) algorithm \cite{gugli:2009} is a probabilistic tool capable to satisfy the 
desiderata and tackle the challenges described in Sect.~\ref{sec:DandC}. 

The BSS algorithm employs the single observed data set (photon image and exposure map) for source 
detection and background estimation. 
Bayesian probability theory (BPT) is the statistical tool used within the BSS technique, supplying a 
general and consistent frame for logical inference. 
Hence, the BSS algorithm takes advantage of all available information over a parameter set, which is 
described by a probability density over the corresponding parameter space. 
For each image pixel $\{ij\}$ two complementary hypotheses are considered
$B_{\rm ij} : d_{\rm ij}=b_{\rm ij}+\epsilon_{\rm ij}$ and $\overline{B}_{\rm ij} : d_{\rm ij}=b_{\rm ij}+s_{\rm ij}+\epsilon_{\rm ij}$. 
Hypothesis $B_{\rm ij}$ specifies that the data $d_{\rm ij}$ consists only of
background counts $b_{\rm ij}$ spoiled with noise
$\epsilon_{\rm ij}$, i.e.~the (statistical) uncertainty associated with the
measurement process. 
Hypothesis $\overline{B}_{\rm ij}$ specifies the case where additional
source intensity $s_{\rm ij}$ contributes to the background. 
Two assumptions are taken into account: No negative values for $s_{\rm ij}$ 
and $b_{\rm ij}$ are allowed and $b_{\rm ij}$ is smoother than $s_{\rm ij}$.
For modelling the structures arising in the background rate of the photon image, 
the thin-plate spline (TPS) is chosen. 
A weighted combination of TPSs centered about each supporting 
point gives the interpolation function that passes through the supporting points exactly while 
minimizing the bending energy. The TPS is not wavering between the supporting points, in opposite to 
cubic-splines, and is steady along the 
field, also where steep gradients in the data occurs, as at the field edge. 
The background amplitude, instead, consists of the TPS multiplied with the exposure map. 
Another important aspect of the technique is the likelihood for the mixture models, that arises 
applying BPT with the mixture model technique. 
The Bayesian mixture model is composed of two parts: a 
Poisson likelihood probability density function (pdf) 
$p(d_{\rm ij} \mid B_{\rm ij},b_{\rm ij})$, i.e.~$b_{\rm ij}$ contribution only, and a 
marginal Poisson likelihood pdf $p(d_{\rm ij} \mid \overline{B}_{\rm ij},b_{\rm ij},\gamma)$, 
i.e.~$b_{\rm ij}$ plus $s_{\rm ij}$ components, where $s_{\rm ij}$ is marginalized and a parameter $\gamma$ is introduced.   
According to BPT, prior pdfs have to be considered for both complementary hypotheses and for 
the source intensity parameter. 
The prior probability for the two complementary hypotheses, i.e.~to have
background only or additional source signal in a pixel or pixel cell\footnote{We define 
pixels as the image finest resolution limited by instrumental design, while we define pixel cell a 
group of correlated neighboring pixels.}, is chosen to be $\beta$ and $1-\beta$. 
Two prior pdfs over the source signal have been considered and tested:
(1) an exponential prior pdf, 
(2) an inverse-$\Gamma$ function prior pdf. 
Both prior pdfs of the source signal introduce the parameter $\gamma$. 
The likelihood for the mixture models results to be:
\begin{eqnarray}
\nonumber
p(D \mid b, \beta, \gamma) = 
\prod_{ij} [\beta \cdot p(d_{\rm ij} \mid B_{\rm ij},b_{\rm ij}) + (1-\beta)
\cdot p(d_{\rm ij} \mid \overline{B}_{\rm ij},b_{\rm ij},\gamma)];\\ \hspace{0.2cm}
D=\{d_{\rm ij}\}, \hspace{0.2cm} b=\{b_{\rm ij}\}.
\label{LMM_app}
\end{eqnarray} 
Equation \ref{LMM_app} allows one to separate background and sources, considering all pixels for 
the background spline estimation, even those containing additional source contribution. It 
expresses our ignorance about the presence of background only or an additional source contribution 
in a certain pixel or pixel cell. This allows us to evaluate the posterior distribution over the 
background, $p(b|D)$, and the probability of having source contributions in pixels and pixel cells, 
$p(\overline{B}_{\rm ij}|d_{\rm ij})$ .\\
$p(b|D)$ is the product of eq.~\ref{LMM_app} and the prior pdf $p(b)$, 
that is chosen constant for positive values of $b$ and null elsewhere. 
The maximum of the posterior pdf with respect to $b$ gives an estimate of the background (amplitude) map. 
The background estimate is provided by the Gaussian approximation, where  
the Hessian matrix is used to extract the uncertainties of the background for each image pixel.
Note that in eq.~\ref{LMM_app} the model parameters $\gamma$ and $\beta$ appear.  
The values of $\gamma$ and $\beta$ and their uncertainties are estimated from the 
marginal posterior pdf $p(\beta,\gamma|D)$, under the assumption of the Laplace approximation.\\ 
The probability $p(\overline{B}_{\rm ij}|d_{\rm ij})$ for each pixel and pixel cells is 
approximated taking into account the optimal values of the background amplitude and the model parameters.  
$p(\overline{B}_{\rm ij}|d_{\rm ij})$ includes the Bayes factor. $p(\overline{B}_{\rm ij}|d_{\rm ij})$ estimated
 for varying correlation lengths of pixels give rise to the multiresolution analysis.   
The multiresolution analysis has the aim to analyze statistically source structures at 
multiple scales. The scales are the correlation lengths employed to create pixel cells. 
Source probability maps (SPMs) are created at different scales. SPMs allow one to separate point-like 
from extended sources. 
Note that background and sources are represented in the mixture components, ensuring no need 
of background subtraction for source detection. Furthermore, Poisson statistics is preserved also in 
the multiresolution analysis.\\
The BSS algorithm allows also for a multiband analysis. 
When multiband images are available, the information contained in each image
  can be statistically combined in order to extend the detection limit of the
  data. Conclusive posterior pdfs are provided for detected sources from combined energy bands.\\  

\begin{figure}
\includegraphics[width=0.41\linewidth,angle=90]{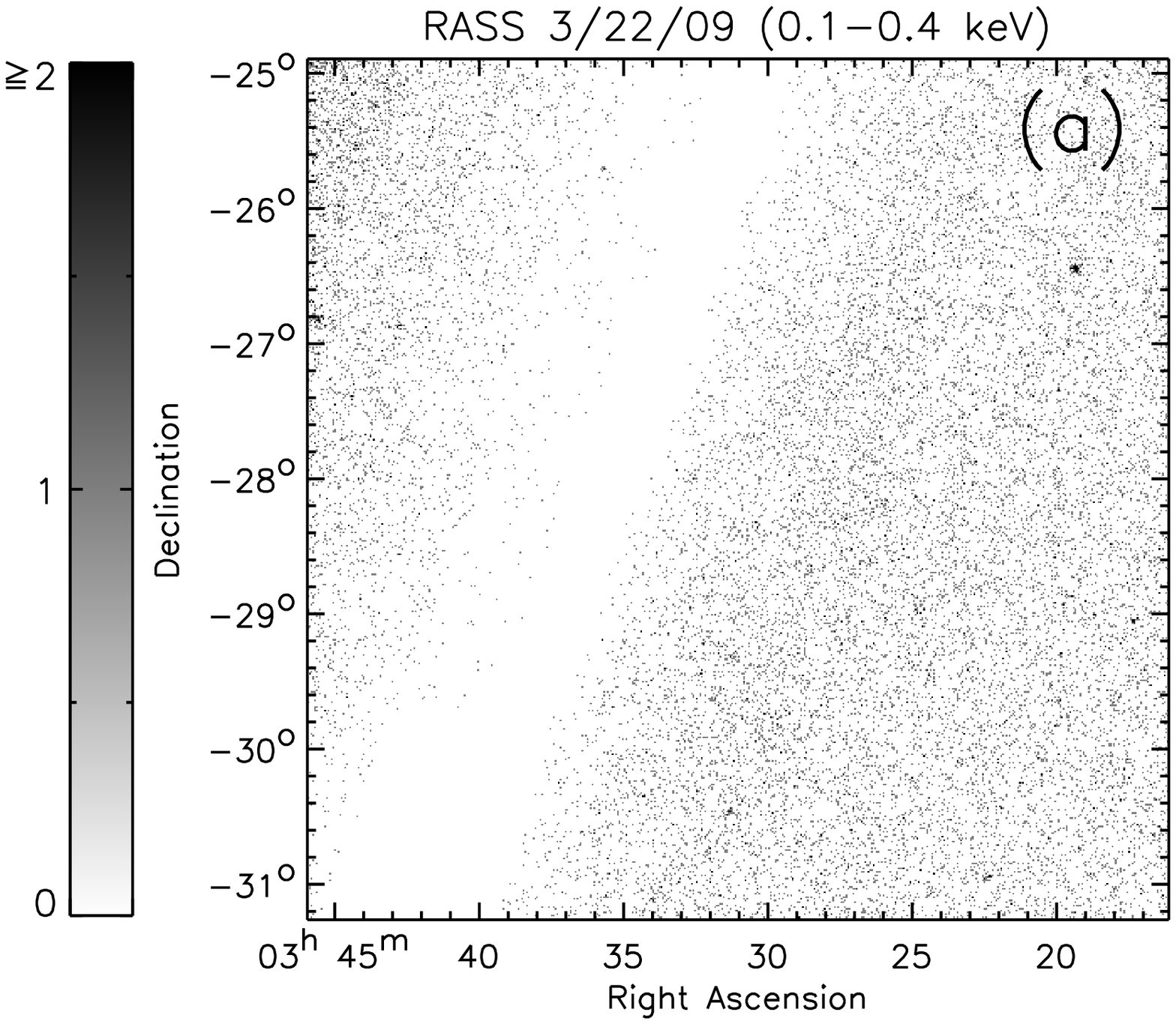}\hspace{\fill}\includegraphics[width=0.4\linewidth,angle=90]{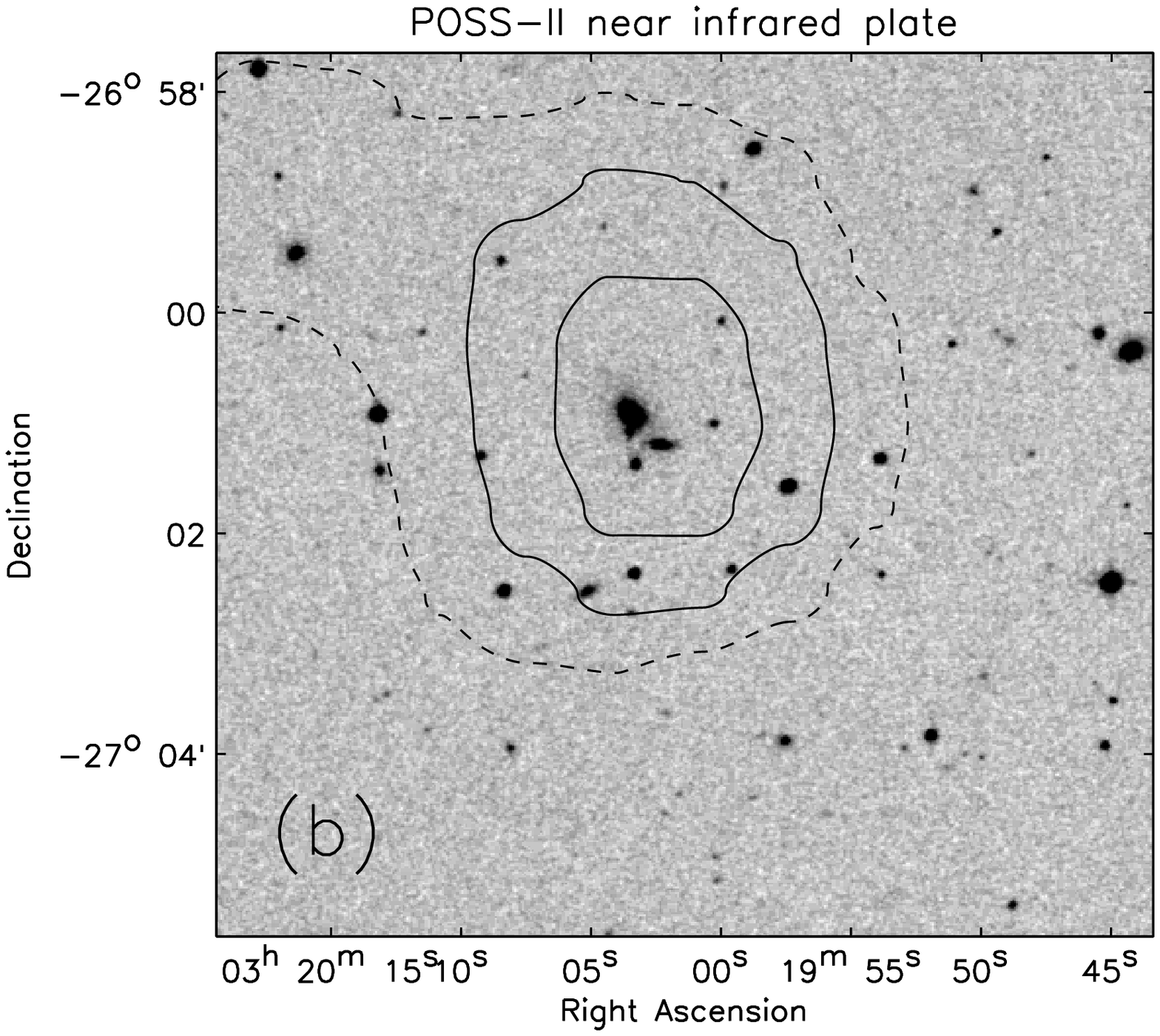}
\caption{Discovery of a cluster of galaxies and confirmed with optical sky surveys. Panel {\bf a}: soft band image of the \emph{ROSAT} All-Sky survey (RASS) field RS932209n00. The image accounts for photon count/pixel in the range $0-9$. Panel {\bf b}: POSS-II I plate with superposed \emph{X}-ray contours from RASS field RS932209n00 (panel {\bf a}) corresponding to $2$, $3$ and $4\sigma$ above the local \emph{X}-ray background. This cluster of galaxies is known in the optical part of the electromagnetic 
spectrum as ACO S 340. } 
\label{disco_clu}
\end{figure}

The BSS algorithm is a general, powerful and flexible Bayesian technique for background and 
source separation. 
The technique is general since it is applicable to astronomical images coming from any count detector. 
The aim of providing more reliable results, with respect to previous techniques, for faint and extended 
sources is achieved: See, as example, in Fig.~\ref{disco_clu} the BSS detection of a new \emph{X}-ray 
cluster of galaxies.  
The technique is flexible, because  it can easily be extended to other statistics and astronomical problems 
in image analysis.  

\acknowledgement{
The first author would like to thank H.~B\"ohringer, 
H.~Brunner and K.~Dennerl (Max-Planck-Institut f\"ur extraterrestrische Physik, Germany), V.~Mainieri and P.~Rosati (European Southern Observatory, Germany) and P.~Tozzi 
(INAF Osservatorio Astronomico di Trieste, Italy) for contributing to this work. 
}

\end{document}